%% file: MinCost-conf-ver4-5.tex
\documentclass[10pt, conference, letterpaper]{IEEEtran}
\IEEEoverridecommandlockouts
\usepackage{cite}
\usepackage{amsmath,amssymb,amsfonts}
\usepackage{algorithmic,algorithm}
\usepackage{graphicx}
\usepackage{caption}
\usepackage{subcaption}

\usepackage{cleveref}
\usepackage{textcomp}
\usepackage{xcolor}
\usepackage{bbm}
\usepackage{tikz}
\usepackage{color,soul}
\usetikzlibrary{patterns}

\newtheorem{theorem}{Theorem}
\newtheorem{lemma}{Lemma}

\newtheorem{definition}{Definition}

\newtheorem{remark}{Remark}

\def\BibTeX{{\rm B\kern-.05em{\sc i\kern-.025em b}\kern-.08em
    T\kern-.1667em\lower.7ex\hbox{E}\kern-.125emX}}
\begin{document}

\title{Minimizing the Sum of Age of Information and Transmission Cost under Stochastic Arrival Model                       
\thanks{We acknowledge support of the Department of Atomic Energy, Government of India, under project no. RTI4001.}
}

\author{\IEEEauthorblockN{Kumar Saurav}
\IEEEauthorblockA{\textit{School of Technology and Computer Science} \\
\textit{Tata Institute of Fundamental Research}\\
Mumbai, India. \\
kumar.saurav@tifr.res.in}
\and
\IEEEauthorblockN{Rahul Vaze}
\IEEEauthorblockA{\textit{School of Technology and Computer Science} \\
\textit{Tata Institute of Fundamental Research}\\
Mumbai, India. \\
rahul.vaze@gmail.com}
}
\vspace{-1in}
\input{input.tex}

\maketitle
\vspace{-0.1in}
\begin{abstract}
We consider a node-monitor pair, where updates are generated stochastically (according to a known distribution) at the node that it wishes to send to the monitor. The node is assumed to incur a fixed cost for each transmission, and the objective of the node is to find the update instants so as to minimize a linear combination of AoI of information and average transmission cost. 
First, we consider the Poisson arrivals case, where updates have an exponential inter-arrival time for which we derive an explicit optimal online policy. 
Next, for arbitrary distributions of inter-arrival time of updates, we propose a simple randomized algorithm that transmits any newly arrived update with a fixed probability (that depends on the distribution) or never transmits that update. The competitive ratio of the proposed algorithm is shown to be a function of the variance and the mean of the inter-arrival time distribution. For some of the commonly considered distributions such as exponential, uniform, and Rayleigh, the competitive ratio bound is shown to be 2.
\end{abstract}


\input{IntroductionKumar.tex}

\section{System Model}
\label{sec:sysModel}
Consider a node, where updates (henceforth, packets) are generated stochastically, with inter-generation time $X$ distributed according to a known distribution $\cD$. 
If the node 
transmits an update to the monitor, it is received instantaneously, without any delay.
At any time $t\ge0$, AoI of the monitor is $\Delta(t)=t-\lambda(t)$, where $\lambda(t)$ denotes the generation time of the latest packet of the node that has also been received by the monitor until time $t$. Therefore, average AoI $\Delta_{av}(t)$ of a node at time $t$ is 
\begin{align}\label{eq:avAge}
\Delta_{av}(t)=\frac{1}{t}\int_{0}^{t}\Delta(i) di.
\end{align}
The node incurs a cost of $c$ units ($c\ge 0$) for each transmission. Hence, average transmission cost at time $t$ is given by
\begin{align}\label{eq:avTxCost}
C_{av}(t)=\frac{c}{t}R(t), 
\end{align}
where, $R(t)$ denotes the number of packets transmitted by the node until time $t$. 
The objective is to obtain a causal optimal transmission policy $\pi^*$, 
\begin{equation} \label{eq:objective}
\pi^*=\underset{\pi\in\Pi}{\arg\min}\lim_{t\rightarrow \infty}\rho C_{av}^{\pi}(t)+\Delta_{av}^{\pi}(t),
\end{equation}
where $\rho\ge0$ is a constant and $\Pi$ is the set of all causal transmission policies $\pi$ (that only requires information obtained until time $t$ to decide whether to transmit at time $t$ or not), while $C_{av}^{\pi}(t)$ and $\Delta_{av}^{\pi}(t)$ denotes the average transmission cost and average AoI on following policy $\pi$, respectively.
\begin{remark} \label{remark:rho}
	Any cost function of the form $aC_{av}(t)+b\Delta_{av}(t)$ (where $a,b>0$ are constants) can be expressed as $b(\rho C_{av}(t)+\Delta_{av}(t))$, (for $\rho=a/b$). Therefore, the solution of corresponding optimization problem is similar to \eqref{eq:objective}. 
\end{remark}

First, we consider the exponential distribution for inter-generation time of packets, and derive an optimal transmission policy for \eqref{eq:objective}. Then in Section \ref{sec:general-distribution}, we generalize this to include general distributions inter-generation time of packets.

\section{Exponential Distribution} \label{sec:Opt-Policy}

In this section, we assume that the inter-generation time of packets is exponentially distributed 
with packet generation rate $q>0$. We develop a causal optimal transmission policy that solves \eqref{eq:objective}, by initially finding a suitable subset within $\Pi$ that contains an optimal solution of \eqref{eq:objective}, and then deriving sufficient conditions for a transmission policy (in the subset) to be optimal, and proposing a transmission policy that satisfies those sufficient conditions.

Among the policies in $\Pi$, 
$\Delta_{av}^{\pi}(t)$ is minimum for that policy $\pi$, which 
at any time, transmits the latest among all the available packets \cite{kaul2012status}. To understand this, note that if a packet is received by the monitor at time $t$, the AoI $\Delta(t)$ decreases to $t-\lambda(t)$ (where $\lambda(t)$ is the generation time of the packet received at time $t$). 
Also, $\lambda(t)$ is maximum (and hence, $t-\lambda(t)$ is minimum) if each time the most recent packet is transmitted. 
So, each time if the most recent packet is transmitted, AoI $\Delta(t)$ is minimum. 
Thus, for an optimal transmission policy $\pi^*$, after a new packet is generated at the node, all previously generated packets become obsolete (as $\pi^*$ never transmits it). 
Hence, the optimal transmission policy $\pi^*\in\Pi_{LCFS}$, where $\Pi_{LCFS}\subseteq\Pi$ is the set of transmission policies
which only transmits the packet with latest generation time. 
In fact, consider the set of transmission policies $\Pi_{NB}\subseteq\Pi_{LCFS}$ 
that either transmits a packet immediately after it is generated, or never transmits it at all.  
Theorem \ref{thm:opt-in-NB} shows that the optimal causal transmission policy $\pi^*\in\Pi_{NB}$. So, \eqref{eq:objective} is equivalent to 
\begin{equation} \label{eq:objective-NB}
\pi^*=\underset{\pi\in\Pi_{NB}}{\arg\min}\lim_{t\rightarrow \infty}\rho C_{av}^{\pi}(t)+\Delta_{av}^{\pi}(t).
\end{equation}

\begin{definition} \label{def:frame}
	Let $t_i$ denote the generation time of $i^{th}$ packet. Then the $i^{th}$ frame $f_i=[t_i,t_{i+1})$ is the time-interval between $i^{th}$ and $(i+1)^{th}$ packet generation. Note that $f_i\cap f_j=\phi$, $\forall i\ne j$, and the time axis can be expressed as $\cup_{i\in\bbN}f_i$. 
\end{definition}

\begin{theorem} \label{thm:opt-in-NB} 
	The optimal transmission policy $\pi^*$ that solves the optimization problem \eqref{eq:objective} either transmits a packet immediatly after it is generated, or never transmits it, i.e., $\pi^*\in\Pi_{NB}$.
\end{theorem}
\begin{IEEEproof} 
	\input{proofThm1.tex}
\end{IEEEproof}

\input{fig_age_noDelay}

Figure \ref{fig:age} shows a possible AoI plot for a policy $\pi_{NB}\in\Pi_{NB}$. A packet generated at time $t_i$ is either transmitted immediately, or never transmitted at all. 
Note that when a transmission policy $\pi_{NB}\in\Pi_{NB}$ is followed, AoI varies in cycles, where a cycle is defined as follows. 
\begin{definition} \label{def:cycle}
	A \emph{cycle} is the interval between the generation time of two consecutively transmitted packets. In particular, let $t_a$ and $t_b$ (where, $t_a<t_b$) denote the generation time of two consecutively transmitted packets. Then the interval $[t_a,t_b)$ represents a cycle, 
	and the length (duration) of the cycle is defined to be $t_b-t_a$. 
\end{definition} 
\begin{remark} \label{remark:cycle-vs-frame}
	Note the difference between the definitions of a cycle and a frame (Definition \ref{def:frame}). As shown in Figure \ref{fig:age}, a frame refers to the interval between generation time of two consecutive packets, whereas a cycle refers to the interval between the generation time of two packet that are also transmitted (consecutively). A cycle may consist of multiple frames, because every generated packet may not be transmitted.
\end{remark}
Henceforth, we denote the $i^{th}$ cycle by $S_i$, and its length (time-duration) by $T_i$. 
AoI cost incurred in a cycle $S_i$ is equal to $\int_{S_i}\Delta(t)dt=T_i^2/2$. 
Also, the number of transmissions in a cycle is exactly 1 (transmission occurs at the start of each cycle). Therefore, the transmission cost incurred in each cycle is $\rho c$. Hence, the total cost incurred in the $i^{th}$ cycle $S_i$ is $\rho c + T_i^2/2$.
Therefore, \eqref{eq:objective-NB} can be expressed as follows.
\begin{align} \label{eq:objective-NB-simplified}
\pi^*&=\underset{\pi\in\Pi_{NB}}{\arg\min}\lim_{t\to\infty}\frac{\sum_{i=1}^{n_t}(\rho c+\frac{1}{2}T_i^2)}{\sum_{i=1}^{n_t}T_i},
\end{align}
where $n_t$ denotes the number of cycles up to time $t$.

Now, let $\Pi_{NB}^{st}$ be the set of all stationary policies in $\Pi_{NB}$ such that $\forall \pi\in\Pi_{NB}^{st}$, $\mathbb{E}_{\pi}[T]<\infty$, where $\mathbb{E}_{\pi}[\cdot]$ denotes expectation with respect to policy $\pi$, and $T$ denotes the cycle length (since $T_i's$ $\forall i$ are i.i.d. under a stationary policy, therefore for concise notation, we drop the subscript $i$ to refer to each $T_i$). 
Then in Lemma \ref{lemma:exp-cost-per-cycle}, we simplify \eqref{eq:objective-NB-simplified} using renewal reward theorem. 
\begin{lemma} \label{lemma:exp-cost-per-cycle}
	With probability 1 (due to renewal reward theorem), \eqref{eq:objective-NB-simplified} is equivalent to
	\begin{align} \label{eq:simplify-to-min-cycle-cost}
	\pi^* &= \underset{\pi\in\Pi_{NB}^{st}}{\arg\min}\left\{\frac{Var_{\pi}(T)+2\rho c}{2\mathbb{E}_{\pi}[T]}+\frac{\mathbb{E}_{\pi}[T]}{2}\right\},
	\end{align}
	where $\mathbb{E}_{\pi}[T]$ and $Var_{\pi}(T)$ denotes the mean and variance of $T$ when policy $\pi$ is followed.
\end{lemma}
\begin{IEEEproof}
	For an optimal policy $\pi^*$, there exists a threshold $\Delta_{max}$ (assuming finite $\rho c$) such that 
	a packet generated at time $t$ is always transmitted if $\Delta(t)>\Delta_{max}$ (otherwise, the additional AoI cost will be larger than the cost due to single transmission i.e. $\rho c$). 
	Therefore if $q>0$, then $\mathbb{E}_{\pi^*}[T]\le \infty$ (in fact, $\mathbb{E}_{\pi^*}[T]\le \Delta_{max}+1/q$, where $1/q$ is the expected inter-generation time of packets at the node). 
	So, we restrict our search space to only those policies $\pi$ within 
	$\Pi_{NB}$ for which $\mathbb{E}_{\pi}[T]<\infty$.
	
	Now, let $Y_i=T_i^2/2+\rho c$. 
	Since the inter-generation time of packets follows exponential distribution, therefore, each cycle is independent of all the previous cycles (i.e., transmission decisions and inter-generation time of packets during previous cycles do not affect the present cycle). So, the optimal transmission policy in a cycle should not depend on the index of the cycle (which implies that the optimal transmission policy should not depend on time). Thus, it is sufficient to search for the optimal transmission policy within the class of stationary transmission policies. 
	So, let $\Pi_{NB}^{st}$ be the set of all stationary policies in $\Pi_{NB}$, such that $\forall \pi\in\Pi_{NB}^{st}$, $\mathbb{E}_{\pi}[T]<\infty$. 
	Therefore, for any transmission policy $\pi\in\Pi_{NB}^{st}$, $((T_1,Y_1),(T_2,Y_2),...)$ forms an independent and identically distributed sequence, and 
	the stochastic process $\mathbf{R}=\{R_t:R_t=\sum_{i=1}^{n_t}Y_i, t\in\mathbb{N}\}$ is a \textit{renewal reward process} \cite{ross2014introduction}. From renewal reward theorem,  
	with probability 1, we have $\lim_{t\to\infty}\frac{R_t}{t}=\frac{\mathbb{E}_{\pi}[Y]}{\mathbb{E}_{\pi}[T]}$ (where $Y=T^2/2+\rho c$). Hence, \eqref{eq:objective-NB} can be written as
	\begin{align} 
	\pi^* &= \underset{\pi\in\Pi_{NB}^{st}}{\arg\min}\frac{\mathbb{E}_{\pi}[\frac{1}{2}T^2+\rho c]}{\mathbb{E}_{\pi}[T]}, \nonumber \\
	&= \underset{\pi\in\Pi_{NB}^{st}}{\arg\min}\frac{\frac{1}{2}(Var_{\pi}(T)+\mathbb{E}_{\pi}[T]^2)+\rho c}{\mathbb{E}_{\pi}[T]}, \nonumber \\
	&= \underset{\pi\in\Pi_{NB}^{st}}{\arg\min}\left\{\frac{Var_{\pi}(T)+2\rho c}{2\mathbb{E}_{\pi}[T]}+\frac{\mathbb{E}_{\pi}[T]}{2}\right\}.  \nonumber 
	\end{align}
\end{IEEEproof}

Let $Var^*(T)=\underset{\pi\in\Pi_{NB}^{st}}{min}Var_{\pi}(T)$, (where $Var_{\pi}(T)$ denotes variance of $T$ when policy $\pi$ is followed). Theorem \ref{thm:opt-condition} lists sufficient conditions for any policy $\pi\in\Pi_{NB}$ to be an optimal solution of \eqref{eq:simplify-to-min-cycle-cost}. 
\begin{theorem} \label{thm:opt-condition}
	A stationary policy $\pi\in\Pi_{NB}^{st}$ is an optimal solution $\pi^*$ of 
	the optimization problem \eqref{eq:simplify-to-min-cycle-cost} if 
	\begin{gather} 
	\label{eq:opt-cond-MeanCycleLength}
	\mathbb{E}_{\pi}[T]=\sqrt{Var^*(T)+2\rho c}, \text{ and} \\
	\label{eq:opt-cond-MinVar}
	Var_{\pi}(T)=Var^*(T).
	\end{gather}
\end{theorem}
\begin{IEEEproof}
	Let $\Gamma_{OPT}$ denote the average cost incurred by the optimal algorithm $\pi^*$ \eqref{eq:simplify-to-min-cycle-cost}. Therefore, 
	\begin{align} \label{eq:Gopt}
	\Gamma_{OPT}&=\underset{\pi\in\Pi_{NB}^{st}}{\min}\left\{\frac{Var_{\pi}(T)+2\rho c}{2\mathbb{E}_{\pi}[T]}+\frac{\mathbb{E}_{\pi}[T]}{2}\right\}.
	\end{align}
	Next, we compute a lower bound on $\Gamma_{OPT}$, and find sufficient conditions for a policy $\pi\in\Pi_{NB}^{st}$ to achieve the lower bound. 
	
	For a lower bound on \eqref{eq:Gopt}, we consider $Var_{\pi}(T)$ and $\bbE_{\pi}[T]$ as independent variables, and then use coordinate-wise minimization approach to minimize \eqref{eq:Gopt} (we show in Lemma \ref{lemma:g-min} that coordinate-wise minimization approach gives a lower bound on \eqref{eq:Gopt}). 
	Initially, we minimize \eqref{eq:Gopt} with respect to $Var_{\pi}(T)\in[Var^*(T),\infty)$ (by definition, $Var^*(T)=\underset{\pi\in\Pi_{NB}^{st}}{min}Var_{\pi}(T)$), and find that it is minimized when $Var_{\pi}(T)=Var^*(T)$. So, we substitute $Var_{\pi}(T)=Var^*(T)$ in \eqref{eq:Gopt}, and then calculate $\bbE_{\pi}[T]\in[0,\infty)$ that minimizes \eqref{eq:Gopt}. We find that for $Var_{\pi}(T)=Var^*(T)$, \eqref{eq:Gopt} is minimized for $\bbE_{\pi}[T]=\sqrt{Var^*(T)+2\rho c}$. 
	So, substituting $Var_{\pi}(T)=Var^*(T)$ and $\bbE_{\pi}[T]=\sqrt{Var^*(T)+2\rho c}$ in \eqref{eq:Gopt}, we get the following lower bound on $\Gamma_{OPT}$ \eqref{eq:Gopt}.
	\begin{align} \label{eq:lb-Gopt}
	\Gamma_{OPT} &\ge \sqrt{Var^*(T)+2\rho c},
	\end{align}
	with equality if 
	for a policy $\pi\in\Pi_{NB}^{st}$, $Var_{\pi}(T)=Var^*(T)$, and $\bbE_{\pi}[T]=\sqrt{Var^*(T)+2\rho c}$. 
	So, a policy $\pi\in\Pi_{NB}^{st}$ achieves the lower bound \eqref{eq:lb-Gopt} on $\Gamma_{OPT}$, and hence is an optimal solution of \eqref{eq:simplify-to-min-cycle-cost},  
	if conditions \eqref{eq:opt-cond-MeanCycleLength} and \eqref{eq:opt-cond-MinVar} are satisfied simultaneously. 
\end{IEEEproof}


Although Theorem \ref{thm:opt-condition} provides sufficient conditions \eqref{eq:opt-cond-MeanCycleLength} and \eqref{eq:opt-cond-MinVar} for optimality, the conditions themselves are in terms of an unknown quantity $Var^*(T)$. 
So, Lemma \ref{lemma:var*} provides an explicit expression for $Var^*(T)$. 
\begin{lemma} \label{lemma:var*} 
	$Var^*(T)=1/q^2$, where $q$ is the packet generation rate. 
\end{lemma}
\begin{IEEEproof}
	 See Appendix \ref{appendix:proof-lemma:var*}.
\end{IEEEproof}


In next section, we propose a threshold-based transmission policy to solve \eqref{eq:simplify-to-min-cycle-cost}. 

\subsection{A Threshold Policy for Packet Transmission} \label{sec:Opt-Transmit-Policy}
Let $\tau$ denote time relative to the start time ($\lambda(t)$) of the ongoing cycle. Thus at time $t$, $\tau=t-\lambda(t)$, 
and whenever a packet is transmitted, new cycle starts and $\tau$ is reset to 0.
Now, consider Algorithm \ref{algo:opt-policy}, a threshold policy for packet transmission. In each cycle, it transmits the first packet that is generated at time $\tau> T^{*}_{q,\rho c}=\sqrt{(1/q^2)+2\rho c}-1/q \ge 0$. Theorem \ref{thm:opt-algo} shows that Algorithm \ref{algo:opt-policy} is an optimal causal transmission policy that solves the optimization problem \eqref{eq:simplify-to-min-cycle-cost}. 

\begin{algorithm}
	\caption{Threshold policy $\pi_{ON}^*$ for packet transmission.}
	\label{algo:opt-policy}
\begin{algorithmic}[1]
	\STATE $\tau\leftarrow 0$; \hspace{4ex} // $\tau$ increases linearly with time.
	\LOOP 
	\IF{a packet is generated \AND $\tau> T^{*}_{q,\rho c}$} 
	\STATE transmit the generated packet;
	\STATE $\tau\leftarrow 0$;
	\ELSE
	\STATE wait for next packet;
	\ENDIF
	\ENDLOOP 
\end{algorithmic}
\end{algorithm}

\begin{theorem} \label{thm:opt-algo}
	The transmission policy $\pi_{ON}^*$ given by Algorithm \ref{algo:opt-policy} with threshold 
	\begin{align} \label{eq:threshold}
	T^{*}_{q,\rho c}=\sqrt{(1/q^2)+2\rho c}-1/q \ge 0
	\end{align}
	is an optimal solution of the optimization problem \eqref{eq:simplify-to-min-cycle-cost}.
\end{theorem}
\begin{IEEEproof} 
	Note that the cycle length $T=T_{q,\rho c}^*+X$, where $X$ is the generation time of first packet after $T_{q,\rho c}^*$. Since $X$ is exponentially distributed with mean $\bbE_q[X]=1/q$. Therefore, $\bbE_{\pi_{ON}^*}[T]=T_{q,\rho c}^*+1/q=\sqrt{(1/q^2)+2\rho c}$. 
	Since $T_{q,\rho c}^*$ is a constant, independent of $X$, therefore, $Var_{\pi_{ON}^*}[T]=Var_{\pi_{ON}^*}(T_{q,\rho c}^*)+Var_q(X)=1/q^2$. Note that $Var^*(T)=1/q^2$ from Lemma \ref{lemma:var*}. So, conditions \eqref{eq:opt-cond-MeanCycleLength} and \eqref{eq:opt-cond-MinVar} are satisfied by $\pi_{ON}^*$ (i.e., Algorithm \ref{algo:opt-policy}). Thus we conclude the result using Theorem \ref{thm:opt-condition}.
\end{IEEEproof}

Apart from being an optimal causal transmission policy, additional interesting guarantee can be established for $\pi_{ON}^*$ (Algorithm \ref{algo:opt-policy}) in terms of an offline optimal transmission policy (that knows all the inputs in advance). This ensures that the performance of $\pi_{ON}^*$ cannot be arbitrarily bad in comparision to a transmission policy that is provided with all input information in advance. 
For a causal policy, a popular approach to establish such a guarantee is 
via competitive ratio ($CR$) bound. $CR$ of a causal policy is defined as the ratio of cost incurred by the causal policy to the cost incurred by an offline optimal policy (that knows the inputs in advance, i.e., in present case, the generation time of all the packets), maximized over all inputs. $CR$ close to 1 guarantees that the performance of the proposed causal policy is close to optimal offline policy. 


\begin{theorem} \label{thm:CR-no-delay}
	$\pi_{ON}^*$ (Algorithm \ref{algo:opt-policy}) has a competitive ratio $CR\le\sqrt{2}$. Additionally, $CR\to 1$ if $q^2\rho c\to\infty$.
\end{theorem}
\begin{IEEEproof}
	See Appendix \ref{appendix:proof-thm:CR-no-delay}.
\end{IEEEproof}


\section{General Distribution} \label{sec:general-distribution}
In Section \ref{sec:Opt-Policy}, we assumed that the inter-generation time of packets follow exponential (memoryless) distribution. However, this assumption might be restrictive in practice. In this section, we generalize our assumption on the distribution for the packet inter-generation time $X$ to arbitrary (non-memoryless) distribution $\cD$. 

Note that when $\cD$ is non-memoryless, we cannot claim the optimal causal transmission policy $\pi^*$ to lie in $\Pi_{NB}$. To understand this, note that when a packet is generated, then waiting for some time may reveal extra information about the generation time of next packet. So, after a packet is generated, optimal policy $\pi^*$ may wait for some time to take informed/optimal decision regarding whether to transmit the packet or not. Moreover, if $\pi^*\notin\Pi_{NB}$, then the AoI cost in different cycles (Definition \ref{def:cycle}) will be inter-dependent, because AoI at the start of each cycle will depend on previous cycle. Also, in continuous time setting, AoI at the start of each cycle will be a continuous random variable that further complicates the analysis. 
So, in this section, we propose a simple stationary randomized policy, and using competitive ratio $(CR)$ analysis, establish guarantees on its performance (for any distribution on packet inter-generation time $X$) relative to the optimal offline solution of \eqref{eq:objective}. 

Let $\{X_i\}_{i\in\bbN}$ denote the inter-generation time of packets at the node, where $X_i's$ ($\forall i\in\bbN$) are i.i.d. with probability density function $\bbP_X$ (such that $\bbP_X(x<0)=0$), and $\bbE_{\bbP_X}[X_i]$ and $Var_{\bbP_X}(X)$ are finite. 
Consider Algorithm \ref{algo:distribution-independent-policy} that transmits each packet immediately after generation with probability $p^*=\min\{\mu_X/\sqrt{\rho c},1\}$ (so, a packet is never transmitted with probability $1-p^*$). 
Theorem \ref{thm:dist-ind-algo-bound-cr} provides a competitive ratio for Algorithm  \ref{algo:distribution-independent-policy} with respect to the optimal transmission policy. 

\begin{algorithm}
	\caption{Stationary randomized policy $\pi_{SR}^*$ for packet transmission.}
	\label{algo:distribution-independent-policy}
	\begin{algorithmic}[1]
		\IF{a packet is generated at the node}
			\STATE transmit the generated packet with probability $p^*$;
		\ELSE
		    \STATE  wait for next packet;
		\ENDIF
	\end{algorithmic}
\end{algorithm}

\begin{theorem} \label{thm:dist-ind-algo-bound-cr}
	The stationary randomized transmission policy $\pi_{SR}^*$ given by Algorithm \ref{algo:distribution-independent-policy} with $p^*=\min\{\mu_X/\sqrt{\rho c},1\}$ has a competitive ratio 
    \begin{align} \label{eq:CR-for-SR}
    	CR_{\pi_{SR}^*}\le\max\left\{2,1+\frac{Var_{\bbP_X}(X)}{\mu_X^2}\right\},
    \end{align}
    where $\mu_X=\bbE_{\bbP_X}[X]$. 
\end{theorem}
\begin{IEEEproof}
To prove Theorem \ref{thm:dist-ind-algo-bound-cr}, we follow a two step approach. In step 1, we compute a lower bound on the average cost for an optimal algorithm. In step 2, we show that in the set of policies $\Pi_{SR}$ that transmits each generated packet immediately with probability $p$, there exists a policy with competitive ratio \eqref{eq:CR-for-SR}. Finally, we show that $\pi_{SR}^*$ given by Algorithm \ref{algo:distribution-independent-policy} with $p=p^*=\min\{\mu_X/\sqrt{\rho c},1\}$ is the optimal policy within $\Pi_{SR}$ to conclude Theorem \ref{thm:dist-ind-algo-bound-cr}. Detailed proof is provided in Appendix \ref{appendix:thm:dist-ind-algo-bound-cr}.
\end{IEEEproof}

Although the competitve ratio \eqref{eq:CR-for-SR} depends on the distribution of inter-generation time of packets, it is bounded for several common distributions. Some examples are as follows. 
\begin{paragraph}
	{Exponential Distribution} For exponential distribution, the ratio $Var_{\bbP_X}(X)/\mu_X^2=1$. Therefore, $CR_{\pi_{SR}^*}\le2$.
\end{paragraph}
\begin{paragraph}
	{Uniform distribution} Let the support of the uniform distribution be over interval $[a,b]$ ($0\le a\le b$). Then $\mu_X=(b+a)/2$, and $Var_{\bbP_X}(X)=(b-a)^2/12$. Therefore, $Var_{\bbP_X}(X)/\mu_X^2\le 1/3$. Hence, $CR_{\pi_{SR}^*}\le\max\{2,4/3\}=2$.
\end{paragraph}
\begin{paragraph}
	{Rayleigh Distribution} Let the scale parameter of rayleigh distribution be $\sigma$. Then $\mu_X=\sigma\sqrt{\pi/2}$, and $Var_{\bbP_X}(X)=\sigma^2(4-\pi)/2$. Therefore, $Var_{\bbP_X}(X)/\mu_X^2=(4/\pi)-1<1$. Hence, $CR_{\pi_{SR}^*}\le 2$.
\end{paragraph}

\input{NumericalResults}

\section{Conclusion} \label{sec:conclusion}
In this paper, we considered a node-monitor pair with stochastic packet inter-generation time, and analysed the problem of minimizing the weighted sum of average AoI and average transmission cost. We derived an optimal transmission policy that is of threshold-type, where we explicitly characterized the threshold. We also showed that for arbitrary distributed packet inter-generation time (with known mean), a simple stationary randomized policy has a bounded competitive ratio (in terms of mean and variance of the distribution).

\appendices

\input{appendices}

\bibliographystyle{IEEEtran}
\bibliography{reflist}

\end{document}

%% file: input.tex
\def\onehalf{\frac{1}{2}}
\def\etal{et.\/ al.\/}
\newcommand{\bydef}{\triangleq}
\newcommand{\tr}{{\it{tr}}}
\def\SNR{{\textsf{SNR}}}
\def\bydef{:=}
\def\bba{{\mathbb{a}}}
\def\bbb{{\mathbb{b}}}
\def\bbc{{\mathbb{c}}}
\def\bbd{{\mathbb{d}}}
\def\bbee{{\mathbb{e}}}
\def\bbff{{\mathbb{f}}}
\def\bbg{{\mathbb{g}}}
\def\bbh{{\mathbb{h}}}
\def\bbi{{\mathbb{i}}}
\def\bbj{{\mathbb{j}}}
\def\bbk{{\mathbb{k}}}
\def\bbl{{\mathbb{l}}}
\def\bbm{{\mathbb{m}}}
\def\bbn{{\mathbb{n}}}
\def\bbo{{\mathbb{o}}}
\def\bbp{{\mathbb{p}}}
\def\bbq{{\mathbb{q}}}
\def\bbr{{\mathbb{r}}}
\def\bbs{{\mathbb{s}}}
\def\bbt{{\mathbb{t}}}
\def\bbu{{\mathbb{u}}}
\def\bbv{{\mathbb{v}}}
\def\bbw{{\mathbb{w}}}
\def\bbx{{\mathbb{x}}}
\def\bby{{\mathbb{y}}}
\def\bbz{{\mathbb{z}}}
\def\bb0{{\mathbb{0}}}

\def\bydef{:=}
\def\ba{{\mathbf{a}}}
\def\bb{{\mathbf{b}}}
\def\bc{{\mathbf{c}}}
\def\bd{{\mathbf{d}}}
\def\bee{{\mathbf{e}}}
\def\bff{{\mathbf{f}}}
\def\bg{{\mathbf{g}}}
\def\bh{{\mathbf{h}}}
\def\bi{{\mathbf{i}}}
\def\bj{{\mathbf{j}}}
\def\bk{{\mathbf{k}}}
\def\bl{{\mathbf{l}}}
\def\bm{{\mathbf{m}}}
\def\bn{{\mathbf{n}}}
\def\bo{{\mathbf{o}}}
\def\bp{{\mathbf{p}}}
\def\bq{{\mathbf{q}}}
\def\br{{\mathbf{r}}}
\def\bs{{\mathbf{s}}}
\def\bt{{\mathbf{t}}}
\def\bu{{\mathbf{u}}}
\def\bv{{\mathbf{v}}}
\def\bw{{\mathbf{w}}}
\def\bx{{\mathbf{x}}}
\def\by{{\mathbf{y}}}
\def\bz{{\mathbf{z}}}
\def\b0{{\mathbf{0}}}
\def\opt{\mathsf{OPT}}
\def\on{\mathsf{ON}}
\def\off{\mathsf{OFF}}
\def\bA{{\mathbf{A}}}
\def\bB{{\mathbf{B}}}
\def\bC{{\mathbf{C}}}
\def\bD{{\mathbf{D}}}
\def\bE{{\mathbf{E}}}
\def\bF{{\mathbf{F}}}
\def\bG{{\mathbf{G}}}
\def\bH{{\mathbf{H}}}
\def\bI{{\mathbf{I}}}
\def\bJ{{\mathbf{J}}}
\def\bK{{\mathbf{K}}}
\def\bL{{\mathbf{L}}}
\def\bM{{\mathbf{M}}}
\def\bN{{\mathbf{N}}}
\def\bO{{\mathbf{O}}}
\def\bP{{\mathbf{P}}}
\def\bQ{{\mathbf{Q}}}
\def\bR{{\mathbf{R}}}
\def\bS{{\mathbf{S}}}
\def\bT{{\mathbf{T}}}
\def\bU{{\mathbf{U}}}
\def\bV{{\mathbf{V}}}
\def\bW{{\mathbf{W}}}
\def\bX{{\mathbf{X}}}
\def\bY{{\mathbf{Y}}}
\def\bZ{{\mathbf{Z}}}
\def\b1{{\mathbf{1}}}

\def\bbA{{\mathbb{A}}}
\def\bbB{{\mathbb{B}}}
\def\bbC{{\mathbb{C}}}
\def\bbD{{\mathbb{D}}}
\def\bbE{{\mathbb{E}}}
\def\bbF{{\mathbb{F}}}
\def\bbG{{\mathbb{G}}}
\def\bbH{{\mathbb{H}}}
\def\bbI{{\mathbb{I}}}
\def\bbJ{{\mathbb{J}}}
\def\bbK{{\mathbb{K}}}
\def\bbL{{\mathbb{L}}}
\def\bbM{{\mathbb{M}}}
\def\bbN{{\mathbb{N}}}
\def\bbO{{\mathbb{O}}}
\def\bbP{{\mathbb{P}}}
\def\bbQ{{\mathbb{Q}}}
\def\bbR{{\mathbb{R}}}
\def\bbS{{\mathbb{S}}}
\def\bbT{{\mathbb{T}}}
\def\bbU{{\mathbb{U}}}
\def\bbV{{\mathbb{V}}}
\def\bbW{{\mathbb{W}}}
\def\bbX{{\mathbb{X}}}
\def\bbY{{\mathbb{Y}}}
\def\bbZ{{\mathbb{Z}}}

\def\cA{\mathcal{A}}
\def\cB{\mathcal{B}}
\def\cC{\mathcal{C}}
\def\cD{\mathcal{D}}
\def\cE{\mathcal{E}}
\def\cF{\mathcal{F}}
\def\cG{\mathcal{G}}
\def\cH{\mathcal{H}}
\def\cI{\mathcal{I}}
\def\cJ{\mathcal{J}}
\def\cK{\mathcal{K}}
\def\cL{\mathcal{L}}
\def\cM{\mathcal{M}}
\def\cN{\mathcal{N}}
\def\cO{\mathcal{O}}
\def\cP{\mathcal{P}}
\def\cQ{\mathcal{Q}}
\def\cR{\mathcal{R}}
\def\cS{\mathcal{S}}
\def\cT{\mathcal{T}}
\def\cU{\mathcal{U}}
\def\cV{\mathcal{V}}
\def\cW{\mathcal{W}}
\def\cX{\mathcal{X}}
\def\cY{\mathcal{Y}}
\def\cZ{\mathcal{Z}}

\def\sfA{\mathsf{A}}
\def\sfB{\mathsf{B}}
\def\sfC{\mathsf{C}}
\def\sfD{\mathsf{D}}
\def\sfE{\mathsf{E}}
\def\sfF{\mathsf{F}}
\def\sfG{\mathsf{G}}
\def\sfH{\mathsf{H}}
\def\sfI{\mathsf{I}}
\def\sfJ{\mathsf{J}}
\def\sfK{\mathsf{K}}
\def\sfL{\mathsf{L}}
\def\sfM{\mathsf{M}}
\def\sfN{\mathsf{N}}
\def\sfO{\mathsf{O}}
\def\sfP{\mathsf{P}}
\def\sfQ{\mathsf{Q}}
\def\sfR{\mathsf{R}}
\def\sfS{\mathsf{S}}
\def\sfT{\mathsf{T}}
\def\sfU{\mathsf{U}}
\def\sfV{\mathsf{V}}
\def\sfW{\mathsf{W}}
\def\sfX{\mathsf{X}}
\def\sfY{\mathsf{Y}}
\def\sfZ{\mathsf{Z}}

\def\bydef{:=}
\def\sfa{{\mathsf{a}}}
\def\sfb{{\mathsf{b}}}
\def\sfc{{\mathsf{c}}}
\def\sfd{{\mathsf{d}}}
\def\sfee{{\mathsf{e}}}
\def\sfff{{\mathsf{f}}}
\def\sfg{{\mathsf{g}}}
\def\sfh{{\mathsf{h}}}
\def\sfi{{\mathsf{i}}}
\def\sfj{{\mathsf{j}}}
\def\sfk{{\mathsf{k}}}
\def\sfl{{\mathsf{l}}}
\def\sfm{{\mathsf{m}}}
\def\sfn{{\mathsf{n}}}
\def\sfo{{\mathsf{o}}}
\def\sfp{{\mathsf{p}}}
\def\sfq{{\mathsf{q}}}
\def\sfr{{\mathsf{r}}}
\def\sfs{{\mathsf{s}}}
\def\sft{{\mathsf{t}}}
\def\sfu{{\mathsf{u}}}
\def\sfv{{\mathsf{v}}}
\def\sfw{{\mathsf{w}}}
\def\sfx{{\mathsf{x}}}
\def\sfy{{\mathsf{y}}}
\def\sfz{{\mathsf{z}}}
\def\sf0{{\mathsf{0}}}

\def\Nt{{N_t}}
\def\Nr{{N_r}}
\def\Ne{{N_e}}
\def\Ns{{N_s}}
\def\Es{{E_s}}
\def\No{{N_o}}
\def\sinc{\mathrm{sinc}}
\def\dmin{d^2_{\mathrm{min}}}
\def\vec{\mathrm{vec}~}
\def\kron{\otimes}
\def\Pe{{P_e}}
\newcommand{\expeq}{\stackrel{.}{=}}
\newcommand{\expg}{\stackrel{.}{\ge}}
\newcommand{\expl}{\stackrel{.}{\le}}
\def\SIR{{\mathsf{SIR}}}

\def\nn{\nonumber}

%% file: IntroductionKumar.tex
\section{Introduction} 
Rapid growth in mobile connectivity and anywhere computing has led to a significant growth in real-time applications of Internet-of-Things (IoT) and Cyber-Physical Systems (CPS). Many of the applications in these paradigms  critically require that fresh status updates are regularly received by the controller, e.g. in health care, delivery apps \cite{myTrackee2018top,boulos2012real,gholamhosseini2019hospital} etc. 
To formally model and capture the concept of freshness of information at the monitor/controller, the metric of {\it age of information} (AoI) \cite{kaul2012real} has been introduced recently, where instantaneous age at any time  is defined as the difference between the current time and the
generation time of the last update that has been successfully received. The AoI is the average of the instantaneous age.

One aspect that is generally neglected when considering AoI optimization is that to transmit an update, a node requires energy, owing to transmission and computation costs. In this paper, we model the energy cost explicitly, and consider a scheduling problem, where the objective function is a linear 
combination of the AoI and the average transmission cost (energy consumed). 

We consider that updates are generated stochastically at the node, with a known inter-generation time distribution. Moreover, to keep the model simple, we assume that each update if sent by the node to the monitor, is received instantaneously, with no delay. 
With transmission cost, clearly, the node cannot transmit all the updates to the monitor. Thus, at each time instant, given the set of outstanding updates that have been generated at the node after the last update was received by the monitor, the decision variable is whether to transmit the most recent outstanding update  or wait for the next update to be generated, given the current age of the monitor, and inter-generation time distribution. 

\subsection{Related Work}
There are primarily two models that are studied with AoI, i) stochastic generation model and ii) generate at will model. We briefly summarize related work in both these directions.
\subsubsection{Stochastic Arrival Model}
The initial work on AoI considered a stochastic model \cite{kaul2012real,kaul2012status,costa2014age}, where the system is modelled as a M/M/1 queue, with inter-generation time of updates and service time (delay seen by each transmission) as exponentially distributed, and found its AoI. In \cite{moltafet2019closed}, a multi-source M/G/1 queueing model (where service time follows general distribution) is considered, and a closed-form expression for the AoI is derived. 
More challenging questions, have been studied more recently, e.g., 
\cite{kavitha2018controlling} considered the problem of identifying the packets that should be transmitted in order to minimize the AoI. 
Further, in context of  energy harvesting nodes, \cite{wu2017optimal, zhou2019transmit} considered AoI minimization problem with stochastic packet generation, subject to energy causality constraints. 
For a more comprehensive review of work of the stochastic arrival model, we refer to \cite{yates2020age}.
\subsubsection{Generate At Will Model}
The generate at will model is becoming more popular recently, where each node always has an update to transmit. 
This model is interesting in two settings, a) either there are multiple nodes that want to update and only a subset of them can update simultaneously but there is no transmission delay, or b) there is a single node, but each update experiences a random delay. 
With multiple nodes, there is a large body of work \cite{bhat2020throughput, kadota2018optimizing, hsu2017scheduling, tripathi2017age, klugel2019aoi, ayan2020optimal},  
when at most one node can transmit at any time without any delay but where each transmission is successful with some probability. With multiple nodes, mostly scheduling algorithms with bounded gap ($2$-competitive) from the optimal algorithms have been derived. For a single node case, when each update experiences a random delay, \cite{sun2017update,sun2017remote} showed that no wait policy (update as soon as the previous update is delivered) is not optimal, and characterized the optimal policy depending on the distribution of the delay. 
Recently, several works \cite{chen2019age,kadota2018scheduling,feng2018minimizing} considered minimizing average AoI with network-related constraints like interference, transmission delay, channel reliability, etc., while \cite{kadota2018optimizing,bhat2020throughput,bastopcu2019age,bastopcu2020partial,talak2019age} considered average AoI minimization problem with other performance metrics like throughput, distortion, delay, etc.  
For more details on prior work we refer the reader to \cite{AoIbook}.

In both the stochastic and the generate at will model, 
the actual cost of transmission can be significant, such as in an IoT setting, where devices are small and have limited energy \cite{fountoulakis2020optimal, nived,xu2020aoi,tseng2019online}. 
The problem of minimizing the linear sum of sampling and transmission cost in a multiple-node system with generate at will model is analyzed in  \cite{fountoulakis2020optimal}, subject to meeting average AoI constraints, and an upper bound on the objective function is derived. 
In \cite{nived}, a multi-node system is considered, where nodes can even transmit their updates (with arbitrary inter-generation time) partially, such that 
the linear sum of AoI, transmission cost and distortion is minimized. A greedy algorithm is proposed that is shown to be $2$-competitive. 
In \cite{tseng2019online}, a node is considered that can download fresh updates (immediately) if a neighboring access-point (AP) is available, and decrease its own instantaneous AoI to $0$. 
The goal is to minimize the linear sum of AoI and downloading cost. 
When the time-slots in which a neigboring AP is available is arbitrary, \cite{tseng2019online} proposes a randomized online algorithm that is $e/(e-1)$-competitive.

\subsection{Our Contributions}
In this paper, we consider a basic scheduling problem, where the objective function is a linear 
combination of the AoI and the average transmission cost (energy consumed), and the decision at each time instant is whether to transmit the 
most recent outstanding update or wait for the next update to arrive at the node, given the current age and the inter-generation time distribution.
\begin{itemize}
\item We first consider the update inter-generation time to be exponentially distributed. For this case, we derive an optimal algorithm, that is threshold based, where we explicitly characterize the threshold as well. Typically, for solving such problem e.g. \cite{sun2017update,xu2020aoi}, structural properties of MDPs are exploited, however, in this work, we take a different approach. We derive a  lower bound on the objective function and derive sufficient conditions to achieve that lower bound. Next, we propose a threshold based algorithm, that transmits a newly generated update if the time since the last transmission is above a certain threshold, and show that it satisfies the optimality conditions.
\item Next, we consider the case of general inter-generation time distributions, and consider a stationary randomized policy, that  either transmits a newly generated update with a certain fixed probability or never transmits it at all. In this setting, we consider the metric of competitive ratio, that is defined as ratio of the cost incurred by the stationary randomized policy to the cost incurred by an offline optimal policy that knows the inputs in advance, maximized over all inputs. For the stationary randomized policy, we derive an upper bound on its competitive ratio in terms of the expectation and variance of the update inter-generation time distributions. 
For commonly considered distributions such as exponential, uniform, and Rayleigh, we show that the competitive ratio of the stationary randomized policy is at most $2$.

\end{itemize}

%% file: proofThm1.tex
\input{fig_piVsBer}
As shown in Figure \ref{fig:diff-pi-piNB}, let the time axis be partitioned into frames (Definition \ref{def:frame}).  
Since a policy $\pi\in\Pi_{LCFS}$ only transmits the latest generated packet, in any frame $f_i$, the number of packets transmitted by $\pi$ is at most 1 (either $i^{th}$ packet, generated at the start of frame $f_i$ is transmitted, or no transmission occurs in the frame at all). 
If $\pi$ transmits the $i^{th}$ packet in frame $f_i$ at time $r_i^{\pi}\in[t_i,t_{i+1})$, then in frame $f_i$, $\pi$ incurs a transmission cost equal to $c$, and the AoI cost equal to $\Delta(t_i)(r_i^{\pi}-t_i)+(t_{i+1}-t_i)^2/2$ (i.e., the area under the AoI plot in frame $f_i$, as shown in Figure \ref{fig:diff-pi-piNB}). Hence, the expected cost 
that $\pi$ incurs in frame $f_i$ is 
\begin{align} \label{eq:do-not-wait}
	\bbE_q&[\rho c+ (\Delta(t_i)(r_i^{\pi}-t_i)+(t_{i+1}-t_i)^2/2)|r_i^{\pi}] \nonumber \\
	\stackrel{(a)}{=}&\rho c +\Delta(t_i)\bbE[r_i^{\pi}-t_i|r_i^{\pi}]+\bbE[(t_{i+1}-r_i^{\pi})^2/2|r_i^{\pi}] \nonumber \\
	&+\bbE[(r_i^{\pi}-t_i)^2/2|r_i^{\pi}]+\bbE[(t_{i+1}-r_i^{\pi})(r_i^{\pi}-t_i)|r_i^{\pi}], \nonumber \\
	\stackrel{(b)}{=}&\rho c +\Delta(t_i)(r_i^{\pi}-t_i)+1/q^2+(r_i^{\pi}-t_i)^2/2 \nonumber \\
	&+(r_i^{\pi}-t_i)/q, 	
\end{align}
where we get $(a)$ by substituting $t_{i+1}-t_i=(t_{i+1}-r_i^{\pi})+(r_i^{\pi}-t_i)$, while $(b)$ follows due to memoryless property of exponential distribution (inter-generation time of packets). 
Note that $r_i^{\pi}\in[t_i,t_{i+1})$. Therefore, the expected cost \eqref{eq:do-not-wait} incurred by $\pi$ in frame $f_i$ is more if $r_i^{\pi}>t_i$ (compared to the case when $r_i^{\pi}=t_i$, i.e., when the $i^{th}$ packet is transmitted immediately after it is generated at time $t_i$). 
Also,  $\forall r_i^{\pi}\in[t_i,t_{i+1})$, the AoI at the start of frame $f_{i+1}$ is same (equal to $t_{i+1}-t_i$, as shown in Figure \ref{fig:diff-pi-piNB}).  
Hence, if $i^{th}$ packet is transmitted by a policy $\pi\in\Pi_{LCFS}$, then it is optimal to transmit it at time $t_i$ (the generation time of $i^{th}$ packet). So, an optimal transmission policy $\pi^*\in\Pi_{LCFS}$ must either transmit a packet immediately after it is generated, or never transmit it (otherwise, another policy $\hat{\pi}^*\in\Pi_{LCFS}$ that transmits the same packets as $\pi^*$, but immediately after they are generated, will incur lesser cost than $\pi^*$, which cannot be true because $\pi^*$ is an optimal transmission policy). 

%% file: fig_piVsBer.tex
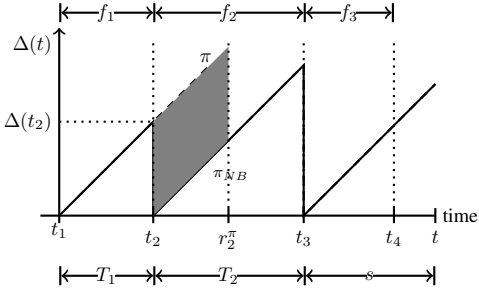
\begin{figure}
	\begin{center}
		\begin{tikzpicture}[thick,scale=1, every node/.style={scale=.75}]
		\draw (0.5,0) to (5.75,0) node[right]{time};
		\draw[->] (0.75,-0.1) to (0.75,2.5) node[below left]{$\Delta(t)$};
		\draw (0.75,0) node[below]{$t_1$} to (2,1.25) to (2,0) to (4,2) to (4,0) to (5.75,1.75); 
		
		\draw[dashed] (1,0.25) to (2.7,1.95) node[above]{$\pi$} to (2.7,0.7) node[below right]{\footnotesize{$\pi_{NB}$}} to (4,2) to (4,0) to (5.75,1.75);
		
		\filldraw[fill=gray, draw=none] (2,1.25) -- (3.0,2.25) -- (3.0,1.0) -- (2,0) -- (2,1.25);
		
		
		
		\draw (2,-0.1) node[below]{$t_2$} to (2,0.1);
		\draw (3.0,-0.1) node[below]{$r_2^{\pi}$} to (3.0,0.1);
		\draw (4,-0.1) node[below]{$t_3$} to (4,0.1);
		\draw (5.2,-0.1) node[below]{$t_4$} to (5.2,0.1);
		\draw (5.75,-0.1) node[below]{$t$} to (5.75,0.1); 
		
		\draw[|<->] (0.75,2.7) -- (2,2.7) node[rectangle,inner sep=-1pt,midway,fill=white]{$f_1$}; 
		\draw[|<->] (2,2.7) -- (4,2.7) node[rectangle,inner sep=-1pt,midway,fill=white]{$f_2$};
		\draw[|<->|] (4,2.7) -- (5.2,2.7) node[rectangle,inner sep=-1pt,midway,fill=white]{$f_3$};
         
         \draw[|<->] (0.75,-0.8) -- (2,-0.8) node[rectangle,inner sep=-1pt,midway,fill=white]{$T_1$}; 
         \draw[|<->] (2,-0.8) -- (4,-0.8) node[rectangle,inner sep=-1pt,midway,fill=white]{$T_2$};
         \draw[|<->|] (4,-0.8) -- (5.75,-0.8) node[rectangle,inner sep=-1pt,midway,fill=white]{$s$};
		
		\draw[dotted] (2,0.1) to (2,2.3);
		\draw[dotted] (3,0.1) to (3,2.3);
		\draw[dotted] (4,0.1) to (4,2.3);
		\draw[dotted] (5.2,0.1) to (5.2,2.3);
		\draw[dotted] (0.75,1.25)  node[left]{$\Delta(t_2)$} to (2,1.25);
		
		\end{tikzpicture}
		\caption{Sample plot of AoI against time when following a policy $\pi\in\Pi_{LCFS}$. When $\pi$ transmits a packet (generated at time $t_i$) at $r_i^{\pi}$, then it incurs an extra AoI cost compared to a policy $\pi_{NB}\in\Pi_{NB}$.\vspace{-4ex}} 
		\label{fig:diff-pi-piNB} 
	\end{center}
\end{figure}

%% file: fig_age_noDelay.tex
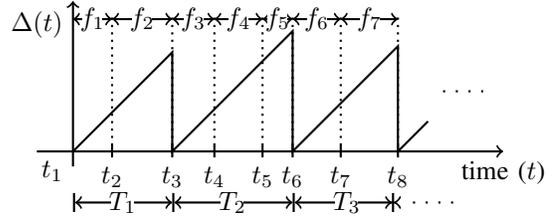
\begin{figure} 
	\begin{center}
		\begin{tikzpicture}[thick,scale=0.8, every node/.style={scale=1}]
		\draw[->] (-0.25,0) to (7.5,0) node[below]{time ($t$)};
		\draw[->] (0.35,-0.25) to (0.35,2.5) node[below left]{$\Delta(t)$};
		\draw (0.35,0) node[below left]{$t_1$} to (2,1.65) to (2,0) to (4,2) to (4,0) to (5.75,1.75) to (5.75,0) to (6.25,0.5); 
		
		
		\draw[loosely dotted] (6.5,1) to (7.3,1); 
		
		\draw (1,-0.1) node[below]{$t_2$} to (1,0.1);
		\draw (2,-0.1) node[below]{$t_3$} to (2,0.1);
		\draw (2.7,-0.1) node[below]{$t_4$} to (2.7,0.1);
		\draw (3.5,-0.1) node[below]{$t_5$} to (3.5,0.1);
		\draw (4,-0.1) node[below]{$t_6$} to (4,0.1);
		\draw (4.8,-0.1) node[below]{$t_7$} to (4.8,0.1);
		\draw (5.75,-0.1) node[below]{$t_8$} to (5.75,0.1); 
		
		\draw[dotted] (1,0.1) to (1,2.3);
		\draw[dotted] (2,0.1) to (2,2.3);
		\draw[dotted] (2.7,0.1) to (2.7,2.3);
		\draw[dotted] (3.5,0.1) to (3.5,2.3);
		\draw[dotted] (4,0.1) to (4,2.3);
		\draw[dotted] (4.8,0.1) to (4.8,2.3);
		\draw[dotted] (5.75,0.1) to (5.75,2.3);
		
		\draw[<->] (0.35,2.2) -- (1,2.2) node[rectangle,inner sep=-1pt,midway,fill=white]{$f_1$}; 
		\draw[<->] (1,2.2) -- (2,2.2) node[rectangle,inner sep=-1pt,midway,fill=white]{$f_2$};
		\draw[<->] (2,2.2) -- (2.7,2.2) node[rectangle,inner sep=-1pt,midway,fill=white]{$f_3$};
		\draw[<->] (2.7,2.2) -- (3.5,2.2) node[rectangle,inner sep=-1pt,midway,fill=white]{$f_4$}; 
		\draw[<->] (3.5,2.2) -- (4,2.2) node[rectangle,inner sep=-1pt,midway,fill=white]{$f_5$};
		\draw[<->] (4,2.2) -- (4.8,2.2) node[rectangle,inner sep=-1pt,midway,fill=white]{$f_6$};
		\draw[<->] (4.8,2.2) -- (5.75,2.2) node[rectangle,inner sep=-1pt,midway,fill=white]{$f_7$};
		
		\draw[|<->] (0.35,-0.85) -- (2,-0.85) node[rectangle,inner sep=-1pt,midway,fill=white]{$T_1$}; 
		\draw[|<->] (2,-0.85) -- (4,-0.85) node[rectangle,inner sep=-1pt,midway,fill=white]{$T_2$};
		\draw[|<->|<] (4,-0.85) -- (5.8,-0.85) node[rectangle,inner sep=-1pt,midway,fill=white]{$T_3$};
		
		\draw[loosely dotted] (6,-0.85) to (6.8,-0.85);
		
		\end{tikzpicture}
		\caption{Sample plot of AoI against time when a transmission policy $\pi_{NB}\in\Pi_{NB}$ is followed. Here, $t_i$ denotes the generation time of $i^{th}$ packet, $f_i$ denotes $i^{th}$ frame, and $T_i$ denotes the length of $i^{th}$ cycle.\vspace{-4ex}} 
		\label{fig:age} 
	\end{center}
\end{figure}

%% file: NumericalResults.tex
\section{Numerical Results} \label{sec:NumericalResults}

In this section, we analyse the performance of the proposed optimal threshold policy $\pi_{ON}^*$ (Algorithm \ref{algo:opt-policy} with threshold $T_{q,\rho c}^*$ \eqref{eq:threshold}) and the proposed stationary randomized policy $\pi_{SR}^*$ (Algorithm \ref{algo:distribution-independent-policy} with $p^*=\min\{\mu_X/\sqrt{\rho c},1\}$) via numerical simulations. 
We let  $\rho=1$, and  consider a sequence of $10000$ packet generations for each simulation. 

\begin{figure}[htbp]
	\centerline{
		\includegraphics[width=0.7\linewidth]{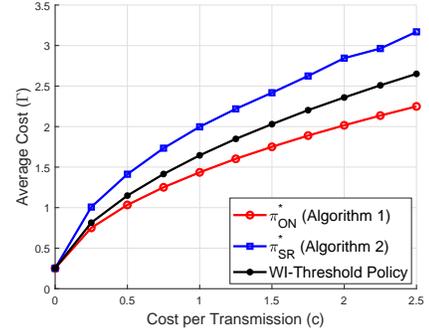}} 
	\caption{Plot of average cost $\Gamma$ of transmission policies w.r.t. cost per transmission.\vspace{-2ex}}
	\label{fig:cost-c}
\end{figure}

Figure \ref{fig:cost-c} shows the plot of average cost $\Gamma$ (sum of average AoI $\Delta_{av}$ and average transmission cost $C_{av}$) with respect to (w.r.t.) cost per transmission ($c$) for $\pi_{ON}^*$ and $\pi_{SR}^*$ when inter-generation time of packets are exponentially distributed with mean $\mu_X=1/q=0.25$. 
When $c$ increases, $\Gamma$ also increases due to increase in $C_{av}$.  However, the threshold $T_{q,\rho c}^*$ \eqref{eq:threshold} for $\pi_{ON}^*$ (Algorithm \ref{algo:opt-policy}) increases with increase in $c$, and the transmission probability $p^*=\min\{\mu_X/\sqrt{\rho c},1\}$ for $\pi_{SR}^*$ (Algorithm \ref{algo:distribution-independent-policy}) decreases with increase in $c$. Thus, when $c$ increases, both $\pi_{ON}^*$ and $\pi_{SR}^*$ decrease the transmission frequency of packets.  
Although $\Gamma$ increases with $c$, the rate of increase in average cost $\Gamma$ is small when $c$ is large. In Figure \ref{fig:cost-c}, also note that the average cost $\Gamma$ for the stationary randomized policy $\pi_{SR}^*$ is less than two (the competitive ratio bound) times the average cost for $\pi_{ON}^*$ as discussed in Section \ref{sec:general-distribution}.  

For comparative analysis, in Figure \ref{fig:cost-c} we also consider WI-threshold policy (WITP) (with  threshold $(\sqrt{0.25+2\rho c /\mu_X}-0.5)\mu_X$), which is a continuous-time equivalent of  
the threshold policy proposed in \cite{kadota2018scheduling} (Proposition 14) for discrete-time model.  
The threshold $T_{q,\rho c}^*$ \eqref{eq:threshold} of the optimal threshold policy $\pi_{ON}^*$ (Algorithm \ref{algo:opt-policy}) depends on $Var_q(X)$, which is not accounted by WITP.  
So, WITP incurs larger average cost $\Gamma$ compared to $\pi_{ON}^*$ as shown in Figure \ref{fig:cost-c}. 

\begin{figure}[htbp]
	\centerline{\includegraphics[width=0.7\linewidth]{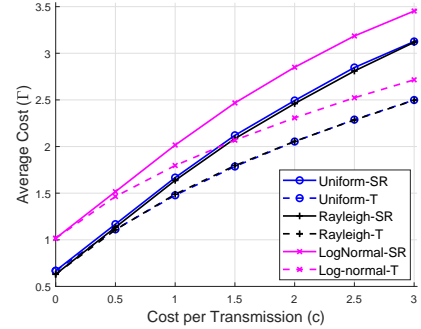}}
	\caption{Plot of average cost $\Gamma$ of transmission policies w.r.t. cost per transmission for different distributions on inter-generation time.\vspace{-4ex}}
	\label{fig:Cost-c-multiDist}
\end{figure}

In Figure \ref{fig:Cost-c-multiDist}, we consider the performance  of $\pi_{SR}^*$ when the distribution on packet inter-generation time is Uniform, Rayleigh and LogNormal, each with mean $\mu_X=1$ and variance $0.33, 0.2732$, and $1$ respectively. 
Also, for each of these distributions, after the sequence of packet generation times were realized, we numerically found a threshold policy $\pi_{T}$ for which the average cost $\Gamma$ is minimum. 
Against each curve, the suffix `-SR' denotes $\pi_{SR}^*$, while the suffix `-T' denotes the threshold policy $\pi_{T}$.  
As shown in Figure \ref{fig:Cost-c-multiDist}, even for the three non-memoryless distributions (for packet inter-generation time) that we considered here, the average cost $\Gamma$ incurred by $\pi_{SR}^*$ is less than two (the competitive ratio \eqref{eq:CR-for-SR} bound for $\pi_{SR}^*$) times the average cost for $\pi_{T}$.

\begin{figure}[htbp]
	\centerline{\includegraphics[width=0.7\linewidth]{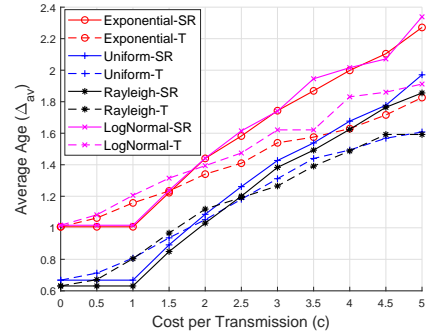}}
    \caption{Variation in average AoI $\Delta_{av}$ w.r.t. cost per transmission for different distributions on inter-generation time.}
	\label{fig:multiDist-avage-c}
\end{figure}

To understand the effect of cost per transmission ($c$) on average AoI $\Delta_{av}$, Figure \ref{fig:multiDist-avage-c} shows the plot of $\Delta_{av}$ against $c$ for Exponential, Uniform, Rayleigh and LogNormal distributions (with mean $\mu_X=1$, and variance 1, 0.33, 0.2732 and 1 respectively) for inter-generation time of packets for both $\pi_{SR}^*$ and $\pi_{T}$. Note that $\pi_{SR}^*$ transmits each packet with probability $p^*=\min\{\mu_X/\sqrt{\rho c},1\}$ (where $\rho$ is a fixed quantity that determines the weightage given to the average transmission cost (w.r.t. average AoI) in the objective function \eqref{eq:objective}; see Remark \ref{remark:rho}). Since $\mu_X=1$ and $\rho=1$, therefore for $c\le 1$, $p^*=1$. Hence when $c\le 1$, $\pi_{SR}^*$ transmits all the packet. So, $\Delta_{av}^{\pi_{SR}^*}$ (average AoI for policy $\pi_{SR}^*$) is constant, and lower than $\Delta_{av}^{\pi_{T}}$ (average AoI for the numerically computed threshold policy $\pi_{T}$) when $c\le 1$. However when $c>1$, $p^*$ decreases with increase in $c$, thereby increasing $\Delta_{av}^{\pi_{SR}^*}$, that ultimately exceeds $\Delta_{av}^{\pi_{T}}$. 

\begin{figure}[htbp]
	\centerline{\includegraphics[width=0.7\linewidth]{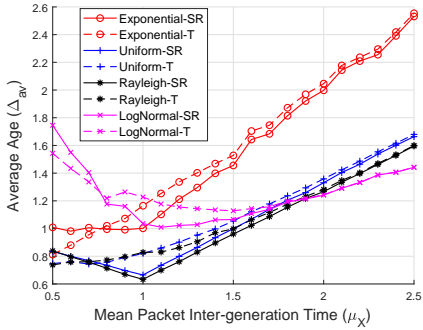}}
	\caption{Variation in average AoI $\Delta_{av}$ w.r.t. mean inter-generation time $\mu_X$ of packets.}
	\label{fig:multiDist-avage-meanX}
\end{figure}

A similar phenomenon is observed in Figure \ref{fig:multiDist-avage-meanX} that shows the plot of $\Delta_{av}^{\pi_{SR}^*}$ (average AoI for policy $\pi_{SR}^*$) w.r.t. $\mu_X$ when $c=1$. In general, it is expected that $\Delta_{av}^{\pi_{SR}^*}$ should increase monotonically with $\mu_X$ (because packets are generated less frequently when $\mu_X$ is large). However, $p^*=\min\{\mu_X/\sqrt{\rho c},1\}$ increases linearly with $\mu_X\in[0,1]$, while $p^*=1$ (constant) when $\mu_X\ge 1$. So, when $\mu_X\in[0,1]$, node transmits more frequently when $\mu_X$ is close to 1. Therefore, $\Delta_{av}^{\pi_{SR}^*}$ either decreases, or increases very slowly. But when $\mu_X\ge 1$, $p^*$ remains constant, and because packets are generated less frequently when $\mu_X$ increases, therefore, $\Delta_{av}^{\pi_{SR}^*}$ increases with increase in $\mu_X\ge 1$.

%% file: appendices.tex
\section{} 
\label{appendix:lemma:g-min}
\begin{lemma} \label{lemma:g-min}
	Let function $g:[d_1,\infty)\times[d_2,\infty)\to \mathbb{R}$ (where $d_1,d_2\in[0,\infty)$ are constants) be defined as $g(x,y)=\frac{y+2\rho c}{x}+x$. 
	Then $g(x,y)\ge g(\max\{d_1,\sqrt{d_2+2\rho c}\},d_2)$. 
\end{lemma}
\begin{IEEEproof}
	Note that $g(x,y)$ is linear in $y$ with minimum at $y=d_2$. Also, for a fixed $y$, $\frac{\partial^2 g(x,y)}{\partial x^2}\ge 0$, $\forall x$. So, $g(x,y)$ is strictly convex in $x$. Moreover, $\frac{\partial g(x,y)}{\partial x}= 0$ at $x=\sqrt{y+2\rho c}$. Hence, for a given $y$, $g(x,y)$ is minimum at $x=\max\{d_1,\sqrt{y+2\rho c}\}$. Therefore, $g(x,y)$ is minimum at $(x,y)=(\max\{d_1,\sqrt{d_2+2\rho c}\},d_2)$, thus proving Lemma \ref{lemma:g-min}.
\end{IEEEproof}

\section{Proof of Lemma \ref{lemma:var*}} \label{appendix:proof-lemma:var*}
A policy $\pi\in\Pi_{NB}^{st}$ either transmits a packet immediately after generation, or never transmits it at all.  
So, cycle length $T$ is equal to the generation time of the transmitted packet (relative to start time of the ongoing cycle). Let $M \ge 1$ denote the index of transmitted packet in a cycle ($M$ may be a random quantity depending on the policy $\pi$). Therefore, $T=\sum_{i=1}^{M-1}X_i+X_M$, where $X_i$ is the inter-generation time of $(i-1)^{th}$ and $i^{th}$ packet. 
Since packet inter-generation time $X$ is i.i.d. exponentially distributed, at any time in a cycle,  next packet is generated after $X$ time units irrespective of the packets generated in the past. So, $X_M$ is independent of $\sum_{i=1}^{M-1}X_i$, irrespective of choice of $M$. 
Hence, $Var_{\pi}(T)=Var_{\pi,q}(\sum_{i=1}^{M-1}X_i)+Var_q(X_M)\ge Var_q(X_M)=1/q^2$.
Also, choosing $M=1$, $Var_{\pi}(T)=1/q^2$ and thus, $Var^*(T)=\underset{\pi\in\Pi_{NB}}{\arg\min} \ \ Var_{\pi}(T)=1/q^2$.

\section{Proof of Theorem \ref{thm:CR-no-delay}} \label{appendix:proof-thm:CR-no-delay}
Let $\pi_{ON}^*$ denote Algorithm \ref{algo:opt-policy} (an optimal online/causal transmission policy) and $\pi_{OFF}^*$ denote an offline optimal transmission policy that knows the generation time of every packet in advance. Also, let $\Gamma_{\pi_{ON}^*}$ and $\Gamma_{\pi_{OFF}^*}$ be the average cost to node (i.e., $\rho C_{av}(t)+\Delta_{av}(t)$) on following $\pi_{ON}^*$ and $\pi_{OFF}^*$, respectively. 
Therefore, the competitive ratio ($CR$) of $\pi_{ON}^*$ relative to $\pi_{OFF}^*$ is given by $CR=\max\frac{\Gamma_{\pi_{ON}^*}}{\Gamma_{\pi_{OFF}^*}}$, where maximization is over all packet generation sequences. 

Note that $\pi_{ON}^*\in\Pi_{NB}^{st}\subseteq\Pi_{NB}$. Therefore, $\pi_{ON}^*$ either transmits a packet immediately after it is generated, or does not transmit it at all. Also, similar to Theorem \ref{thm:opt-in-NB}, it can be argued that $\pi_{OFF}^*$ either transmits a packet immediately after it is generated, or drops it forever. So, AoI plot for both $\pi_{ON}^*$ and $\pi_{OFF}^*$ is as shown in Figure \ref{fig:age}, and average cost to node is $\lim_{t\to\infty}\frac{\sum_{i=1}^{n_t}(\frac{1}{2}T_i^2+\rho c)}{\sum_{i=1}^{n_t}T_i}$, where $T_i\ge0$ is the length of $i^{th}$ cycle, and $n_t\ge0$ is the number of cycles up to time $t$. Using renewal reward theorem as in the proof of Lemma \ref{lemma:exp-cost-per-cycle}, we get 
\begin{align} \label{eq:gamma-on}
\Gamma_{\pi_{ON}^*}&=\frac{1}{2}\left(\frac{Var_{\pi_{ON}^*}(T)+2\rho c}{\mathbb{E}_{\pi_{ON}^*}[T]}+\mathbb{E}_{\pi_{ON}^*}[T]\right),  \\
\label{eq:gamma-off} 
\Gamma_{\pi_{OFF}^*}&=\frac{1}{2}\left(\frac{Var_{\pi_{OFF}^*}(T)+2\rho c}{\mathbb{E}_{\pi_{OFF}^*}[T]}+\mathbb{E}_{\pi_{OFF}^*}[T]\right).
\end{align} 

Now, substituting  $\mathbb{E}_{\pi_{ON}^*}[T]=\sqrt{(1/q^2)+2\rho c}$ and $Var_{\pi_{ON}^*}(T)=1/q^2$ in \eqref{eq:gamma-on}, we get 
\begin{align}\label{eq:gamma-on-ineq}
	\Gamma_{\pi_{ON}^*} &= \sqrt{(1/q^2)+2\rho c}. 
\end{align}

Also, using Lemma \ref{lemma:g-min} (see Appendix \ref{appendix:lemma:g-min}) we get a lower bound on \eqref{eq:gamma-off}:  
\begin{align} \label{eq:lb-prelim-Gamma-OFF}
\Gamma_{\pi_{OFF}^*}\ge\sqrt{Var_{\pi_{OFF}^*}+2\rho c}.
\end{align}
Since this lower bound is in terms of $Var_{\pi_{OFF}^*}(T)$ that is unknown, so for an explicit lower bound on $\Gamma_{\pi_{OFF}^*}$, we next compute a lower bound on $Var_{\pi_{OFF}^*}(T)$. 
For ease of notation, let $\hat{\mu}$ denote $\bbE_{\pi_{OFF}^*}[T]$. Then by definition of variance, 
$Var_{\pi_{OFF}^*}(T)$ 
\begin{align} \label{eq:var-pioff-express}
&=\int_{0}^{\infty}\mathbb{P}((T-\hat{\mu})^2>x)dx 
\stackrel{(a)}{=}\int_{0}^{\infty}2y\mathbb{P}(|T-\hat{\mu}|>y)dy, \nonumber \\
&=\int_{0}^{\hat{\mu}}2y\mathbb{P}(|T-\hat{\mu}|>y)dy +\int_{\hat{\mu}}^{\infty}2y\mathbb{P}(|T-\hat{\mu}|>y)dy, 
\end{align}
where in $(a)$, we used change of variables (replaced $x$ by $y^2$).
Note that $T$ is equal to the generation time of the transmitted packet (relative to the start time of the cycle). So, $T$ cannot lie in interval $[\hat{\mu}-y, \hat{\mu}+y]$ if no packet is generated in this interval of time (this is sufficient condition, but not necessary). So, $\mathbb{P}(|T-\hat{\mu}|>y)$ is greater than the probability that no packet is generated in the interval $[\hat{\mu}-y, \hat{\mu}+y]$. Since packet inter-generation time follows exponential (memoryless) distribution with parameter $q$, therefore, if $y\le\hat{\mu}$, then $\mathbb{P}(|T-\hat{\mu}|>y)$ is lower bounded by $e^{-q2y}$, otherwise it is lower bounded by $e^{-q(\hat{\mu}+y)}$. 
So, 
from \eqref{eq:var-pioff-express} we get
\begin{align} 
Var_{\pi_{OFF}^*}(T) &\ge\int_{0}^{\hat{\mu}}2ye^{-q2y}dy+\int_{\hat{\mu}}^{\infty}2ye^{-q(\hat{\mu}+y)}dy, \\ \label{eq:var-pioff-lowerbound}
&=\frac{1}{2q^2}+2e^{-2q\hat{\mu}}\left(\frac{\hat{\mu}}{2q}+\frac{3}{4q^2}\right) \stackrel{(a)}{\ge} \frac{1}{2q^2},
\end{align}
where we got $(a)$ by minimizing with respect to $\hat{\mu}$.    
Thus,  from \eqref{eq:lb-prelim-Gamma-OFF} and \eqref{eq:var-pioff-lowerbound} we get
\begin{align}\label{eq:gamma-off-ineq}
\Gamma_{\pi_{OFF}^*}
&\ge \sqrt{(1/2q^2)+2\rho c}.
\end{align}
So, using \eqref{eq:gamma-on-ineq} and \eqref{eq:gamma-off-ineq}, we get the competitive ratio
\begin{align}
CR&= \frac{\sqrt{1+2q^2\rho c}}{\sqrt{0.5+2q^2\rho c}} \le \sqrt{2}. 
\end{align}
Note that if $q^2\rho c$ is large, then $CR$ is close to 1.

\section{Proof of Theorem \ref{thm:dist-ind-algo-bound-cr}} \label{appendix:thm:dist-ind-algo-bound-cr}
We prove Theorem \ref{thm:dist-ind-algo-bound-cr} in two steps. First, we compute a lower bound on the average cost for an optimal algorithm, and then we compute the competitive ratio for $\pi_{SR}^*$ (Algorithm \ref{algo:distribution-independent-policy} with $p^*=\min\{\mu_X/\sqrt{\rho c},1\}$). 
\subsubsection{Lower bound}
For lower bound, consider an offline optimal transmission policy $\pi_{OFF}^*$. Since an offline policy knows the generation time of all the packets in advance, therefore, regardless of the distribution on packet inter-generation time $X$, it must transmit the packet that is to be transmitted, immediately after generation (otherwise, if a packet is stored and transmitted later, then the node incurs extra AoI cost (similarly as shown in Figure \ref{fig:diff-pi-piNB}) without any reduction in transmission cost). Therefore, AoI of the node under an offline optimal policy varies in cycles as shown in Figure \ref{fig:age} (i.e., in each cycle, AoI increases linearly with time, and then instantly drops to 0). Therefore, AoI cost incurred in $i^{th}$ cycle is $T_i^2/2$, where $T_i$ denotes the length of $i^{th}$ cycle. Also, at time $t$, let $s=t-\sum_{i=1}^{R(t)}T_i$ denote the time elapsed since last packet transmission (i.e., the length of ongoing (incomplete) cycle at time $t$ as shown in Figure \ref{fig:diff-pi-piNB}). Therefore, overall AoI cost for $\pi_{OFF}^*$ is $\sum_{i=1}^{R(t)}(T_i^2/2)+s^2/2$. Hence,
\begin{align} \label{eq:lb-age-1}
	\Delta_{av}(t)&=\frac{1}{t}\left[\left(\sum_{i=1}^{R(t)}\frac{T_i^2}{2}\right)+\frac{s^2}{2}\right], \nonumber \\
	&=\frac{1}{2}\left[\frac{R(t)}{t}\frac{1}{R(t)}\sum_{i=1}^{R(t)}T_i^2+\frac{s^2}{t}\right], \nonumber \\
	&\stackrel{(a)}{\ge}\frac{1}{2}\left[\frac{R(t)}{t}\left(\frac{1}{R(t)}\sum_{i=1}^{R(t)}T_i\right)^2+\frac{s^2}{t}\right], \nonumber \\
	&\ge\frac{1}{2}\left[\frac{1}{t}\frac{(t-s)^2}{R(t)}+\frac{s^2}{t}\right],
\end{align}
where in $(a)$, we used Jensen's inequality. Minimizing \eqref{eq:lb-age-1} with respect to $s\in[0,\infty)$, we find that \eqref{eq:lb-age-1} is minimum for $s=t/(1+R(t))$. Thus, substituting $s=t/(1+R(t))$ in \eqref{eq:lb-age-1}, we get
\begin{align} \label{eq:lb-age-2}
\Delta_{av}(t) &\ge \frac{1}{2t}\left[\frac{(t-t/(1+R(t)))^2}{R(t)}+\left(\frac{t}{1+R(t)}\right)^2\right], \nonumber \\
&=\frac{t/2}{1+R(t)}.
\end{align}
A lower bound similar to \eqref{eq:lb-age-2} with an additive term $1/2$ was computed in \cite{kadota2018optimizing} for average AoI for a discrete-time model. Now, since $C_{av}(t)=cR(t)/t$, therefore, substituting $R(t)/t=C_{av}(t)/c$ in \eqref{eq:lb-age-2}, and taking limits as $t\to\infty$ in \eqref{eq:lb-age-2}, we get
\begin{align}\label{eq:lb-age-3}
\lim_{t\to\infty}\Delta_{av}(t)&\ge\lim_{t\to\infty}\frac{1/2}{C_{av}(t)/c+1/t} =\lim_{t\to\infty}\frac{c/2}{C_{av}(t)}.
\end{align}
Therefore, using \eqref{eq:lb-age-3} we get a lower bound on $\Gamma_{\pi_{OFF}^*}$ (where $\Gamma_{\pi_{OFF}^*}$ is the average cost on following $\pi_{OFF}^*$) as follows.
\begin{align} \label{eq:lb-gamma-opt}
\Gamma_{\pi_{OFF}^*}&=\underset{\pi\in\Pi}{\min}\lim_{t\to\infty}\left\{\Delta_{av}^{\pi}(t)+\rho C_{av}^{\pi}(t)\right\}, \nonumber \\
&\stackrel{(a)}{=} \Delta_{av}^{\pi_{OFF}^*}(t)+\rho C_{av}^{\pi_{OFF}^*}(t), \nonumber \\
&\ge \lim_{t\to\infty}\frac{c/2}{C_{av}^{\pi_{OFF}^*}(t)}+\rho C_{av}^{\pi_{OFF}^*}(t), \nonumber \\ 
&\stackrel{(b)}{=} \frac{c/2}{C_{av}^{\pi_{OFF}^*}}+\rho C_{av}^{\pi_{OFF}^*},
\end{align}
where in $(a)$, the policy $\pi_{OFF}^*$ is the optimal offline policy that minimizes $\Delta_{av}(t)+\rho C_{av}(t)$ as $t\to\infty$, while in $(b)$, $C_{av}^{\pi_{OFF}^*}=\lim_{t\to\infty}C_{av}^{\pi_{OFF}^*}(t)$. 

\subsubsection{Competitive Ratio}
Let $\Pi_{SR}$ be the set of transmission policies that transmit each packet immediately after generation with probability $p$, or never transmit the packet. 
Since $\Pi_{SR}\subseteq\Pi_{NB}$, therefore as shown in Section \ref{sec:Opt-Policy}, for any policy $\pi_{SR}\in\Pi_{SR}$, we have
\begin{align} \label{eq:gamma-pisr-1}
\Gamma_{\pi_{SR}}&=\frac{Var_{\pi_{SR}}(T)+2\rho c}{2\mathbb{E}_{\pi_{SR}}[T]}+\frac{\mathbb{E}_{\pi_{SR}}[T]}{2}, \nonumber \\
&=\frac{\bbE_{\pi_{SR}}[T^2]}{2\bbE_{\pi_{SR}}[T]}+\frac{\rho c}{\bbE_{\pi_{SR}}[T]}, \nonumber \\
&=\bbE_{\pi_{SR}}[T]\left(\frac{\bbE_{\pi_{SR}}[T^2]}{2\bbE_{\pi_{SR}}[T]^2}\right)+\frac{\rho c}{\bbE_{\pi_{SR}}[T]}. 
\end{align}
Note that $T_i=\sum_{j=1}^{m_i}X_{ij}$, where $m_i$ is 
the index of the generated packet in $i^{th}$ cycle that is transmitted, and $X_{ij}$ 
is equal to the inter-generation time of $(j-1)^{th}$ and $j^{th}$ generated packet in $i^{th}$ cycle. So, $T_i$ is a sum of random number ($m_i$) of i.i.d. random variables ($X_{ij}$). Also,  $m_i$ is independent of $X_{ij}$ ($\forall j$), because $\pi_{SR}$ transmits each packet with probability $p$ (independent of $X_{ij}$). Hence, using Wald's equation \cite{mckay2019probability},
\begin{align} \label{eq:mean-T}
\bbE_{\pi_{SR}}[T_i]=\bbE_{\pi_{SR}}[m_i]\bbE_{\bbP_X}[X]=\mu_X/p, 
\end{align}
where $\mu_X=\bbE_{\bbP_X}[X]$.
Also, $T_i^2=\sum_{j=1}^{m_i}X_{ij}^2+ \sum_{j=1}^{m_i}\sum_{k=1,k\ne j}^{m_i}X_{ij}X_{ik}$. Therefore, 
\begin{align} \label{eq:mean-T2}
\bbE_{\pi_{SR}}[T_i^2]&\stackrel{(a)}{=}\bbE_{\pi_{SR}}[m_i]\mu_{X^2}+\bbE_{\pi_{SR}}[m_i^2-m_i]\mu_X^2, \nonumber \\
&=\bbE_{\pi_{SR}}[m_i]Var_{\bbP_X}(X)+\bbE_{\pi_{SR}}[m_i^2]\mu_X^2, 
\end{align}
where in $(a)$, $\mu_{X^2}=\bbE_{\bbP_X}[X^2]$, and $\mu_X=\bbE_{\bbP_X}[X]$.
Therefore, using \eqref{eq:mean-T} and \eqref{eq:mean-T2} (and dropping the subscript $i$ for ease of notation), we get
\begin{align} \label{eq:ratio-meanT2-meanT}
\frac{\bbE_{\pi_{SR}}[T^2]}{\bbE_{\pi_{SR}}[T]^2}&=\frac{\bbE_{\pi_{SR}}[m]Var_{\bbP_X}(X)+\bbE_{\pi_{SR}}[m^2]\mu_X^2}{(\bbE_{\pi_{SR}}[m]\mu_X)^2}, \nonumber \\
&=\frac{Var_{\bbP_X}(X)}{\bbE_{\pi_{SR}}[m]\mu_X^2}+\frac{\bbE_{\pi_{SR}}[m^2]}{\bbE_{\pi_{SR}}[m]^2}, \nonumber \\
&\stackrel{(a)}{=}p\frac{Var_{\bbP_X}(X)}{\mu_X^2}+2-p, \nonumber \\
&=2-p\left(1-\frac{Var_{\bbP_X}(X)}{\mu_X^2}\right),
\end{align}
where in $(a)$, we used $\bbE[m]=1/p$, and $\bbE[m^2]=(2-p)/p^2$.
Substituting \eqref{eq:ratio-meanT2-meanT} into \eqref{eq:gamma-pisr-1}, we get
\begin{align} \label{eq:gamma-pisr-2}
\Gamma_{\pi_{SR}}=\frac{\bbE_{\pi_{SR}}[T]}{2}\left(2-p\left(1-\frac{Var_{\bbP_X}(X)}{\mu_X^2}\right)\right)+\frac{\rho c}{\bbE_{\pi_{SR}}[T]}.
\end{align}
Since $\bbE[X_i]<\infty$, therefore for $p>0$, $R_{\pi_{SR}}(t)\to\infty$ as $t\to\infty$ (where $R_{\pi_{SR}}(t)$ denotes number of packets transmitted until time $t$ when policy $\pi_{SR}$ is followed). Also, $T_i$ are i.i.d. for each $i\in\bbN$.
So, using strong law of large numbers, 
\begin{gather} 
\label{eq:avg-to-mean-T}
\lim_{t\rightarrow \infty}\frac{1}{R_{\pi_{SR}}(t)}\sum_{i=1}^{R_{\pi_{SR}}(t)}T_i\stackrel{a.s.}{\rightarrow}\bbE_{\pi_{SR}}[T]. 
\end{gather}
Let $C_{av}^{\pi_{SR}}=\lim_{t\to\infty}C_{av}^{\pi_{SR}}(t)$. Therefore, by definition, 
\begin{align} \label{eq:cav-pisr-1}
C_{av}^{\pi_{SR}}=\lim_{t\to\infty}\frac{cR_{\pi_{SR}}(t)}{t}=\lim_{t\to\infty}\frac{cR_{\pi_{SR}}(t)}{\sum_{i=1}^{R_{\pi_{SR}}(t)}T_i}. 
\end{align}
So, from \eqref{eq:avg-to-mean-T} and \eqref{eq:cav-pisr-1} we have
\begin{align} \label{eq:cav-pisr-2}
C_{av}^{\pi_{SR}}\stackrel{a.s.}{\to}\frac{c}{\bbE_{\pi_{SR}}[T]},
\end{align}
Thus, using \eqref{eq:gamma-pisr-2} and \eqref{eq:cav-pisr-2}, we get
\begin{align} \label{eq:gamma-pisr-3}
\Gamma_{\pi_{SR}}=\frac{c/2}{C_{av}^{\pi_{SR}}}\left(2-p\left(1-\frac{Var_{\bbP_X}(X)}{\mu_X^2}\right)\right)+\rho C_{av}^{\pi_{SR}}.
\end{align}
Note that for policy $\pi_{SR}\in\Pi_{SR}$ with $p=0$, $C_{av}^{\pi_{SR}}=0$, while for a policy $\pi_{SR}\in\Pi_{SR}$ with $p=1$,$C_{av}^{\pi_{SR}}$ is maximum for any given realization of the packet generation process. Hence, there exists a certain value of $p\in[0,1]$ (say, $\hat{p}$), such that for the stationary policy $\pi_{SR,\hat{p}}\in\Pi_{SR}$ with $p=\hat{p}$, $C_{av}^{\pi_{SR,\hat{p}}}=C_{av}^{\pi_{OFF}^*}$ (where $\pi_{OFF}^*$ is the optimal offline transmission policy  used in \eqref{eq:lb-gamma-opt}). 
Therefore, using \eqref{eq:lb-gamma-opt} and \eqref{eq:gamma-pisr-3} we get
\begin{align} \label{eq:ratio-gamma-sr-phat-vs-opt-1}
\frac{\Gamma_{\pi_{SR,\hat{p}}}}{\Gamma_{\pi_{OFF}^*}}&\le\frac{\frac{c/2}{C_{av}^{\pi_{OFF}^*}}\left(2-\hat{p}\left(1-\frac{Var_{\bbP_X}(X)}{\mu_X^2}\right)\right)+\rho C_{av}^{\pi_{OFF}^*}}{\frac{c/2}{C_{av}^{\pi_{OFF}^*}}+\rho C_{av}^{\pi_{OFF}^*}}, \nonumber\\
&=\frac{\left(2-\hat{p}\left(1-\frac{Var_{\bbP_X}(X)}{\mu_X^2}\right)\right)+\frac{2\rho}{c} (C_{av}^{\pi_{OFF}^*})^2}{1+\frac{2\rho}{c} (C_{av}^{\pi_{OFF}^*})^2},
\nonumber\\
&=\frac{\left(2-\hat{p}\left(1-\frac{Var_{\bbP_X}(X)}{\mu_X^2}\right)\right)-1}{1+\frac{2\rho}{c} (C_{av}^{\pi_{OFF}^*})^2}+1, \nonumber \\
&\le 2-\hat{p}\left(1-\frac{Var_{\bbP_X}(X)}{\mu_X^2}\right). 
\end{align}
From \eqref{eq:ratio-gamma-sr-phat-vs-opt-1}, it follows that  
\begin{enumerate}
	\item if $\frac{Var_{\bbP_X}(X)}{\mu_X^2} \le 1$, then $\frac{\Gamma_{\pi_{SR,\hat{p}}}}{\Gamma_{\pi_{OFF}^*}} \le 2$, and 
	\item if $\frac{Var_{\bbP_X}(X)}{\mu_X^2} > 1$, then $\frac{\Gamma_{\pi_{SR,\hat{p}}}}{\Gamma_{\pi_{OFF}^*}}\le 2-(1-\frac{Var_{\bbP_X}(X)}{\mu_X^2})$. 
\end{enumerate}
Therefore, $\frac{\Gamma_{\pi_{SR,\hat{p}}}}{\Gamma_{\pi_{OFF}^*}}\le\max\left\{2,2-\left(1-\frac{Var_{\bbP_X}(X)}{\mu_X^2}\right)\right\}$, i.e., 
\begin{align}
\label{eq:ratio-gamma-sr-vs-opt}
\frac{\Gamma_{\pi_{SR,\hat{p}}}}{\Gamma_{\pi_{OFF}^*}}&\le \max\left\{2,1+\frac{Var_{\bbP_X}(X)}{\mu_X^2}\right\}.
\end{align}
Now, let $\pi_{SR}^*\in\Pi_{SR}$ be the optimal policy (among the policies in $\Pi_{SR}$) with $p=p^*$ such that $\Gamma_{\pi_{SR}^*}\le\Gamma_{\pi_{SR}}$ $\forall \pi_{SR}\in\Pi_{SR}$. Therefore, $\Gamma_{\pi_{SR}^*}\le\Gamma_{\pi_{SR,\hat{p}}}$. Thus, from \eqref{eq:ratio-gamma-sr-vs-opt} we get the competitive ratio for $\pi_{SR}^*$ to be
\begin{align}
	CR_{\pi_{SR}^*}\le\max\left\{2,1+\frac{Var_{\bbP_X}(X)}{\mu_X^2}\right\}.
\end{align}

Next, we find $\pi_{SR}^*$ by computing $p^*$ that minimizes $\Gamma_{\pi_{SR}}$ \eqref{eq:gamma-pisr-2}. Substituting $\bbE_{\pi_{SR}}[T]=\mu_X/p$ from \eqref{eq:mean-T} into \eqref{eq:gamma-pisr-2}, we get $\Gamma_{\pi_{SR}}=\frac{\mu_X}{2p}\left(2-p\left(1-\frac{Var_{\bbP_X}(X)}{\mu_X^2}\right)\right)+p\frac{\rho c}{\mu_X}$, i.e., 
\begin{align} \label{eq:gamma-pisr-4}
\Gamma_{\pi_{SR}} 
&=\frac{\mu_X}{p}+p\frac{\rho c}{\mu_X}-\frac{\mu_X}{2}\left(1-\frac{Var_{\bbP_X}(X)}{\mu_X^2}\right).
\end{align}
Relaxing $p$ to take values in $(0,\infty]$, we find that \eqref{eq:gamma-pisr-4} is minimum at $p=\mu_X/\sqrt{\rho c}$. Since $p^*\in(0,1]$, therefore, if $\mu_X/\sqrt{\rho c}\in(0,1]$, then $p^*=\mu_X/\sqrt{\rho c}$. Otherwise, $p^*=1$ because \eqref{eq:gamma-pisr-4} is convex in $p$, and hence, non-increasing in $p$ in interval $(0,\mu_X/\sqrt{\rho c}]$. Therefore, if $\mu_X/\sqrt{\rho c}>1$, then  for $p\in(0,1]$, $\Gamma_{\pi_{SR}}$ is minimum at $p=1=p^*$.  
Thus collectively, $p^*=\min\{\mu_X/\sqrt{\rho c},1\}$. 